\documentclass[
epj,
nopacs,
]{svjour}
\usepackage{cite}
\usepackage{notoccite}
\usepackage{graphicx}
\usepackage{dcolumn}
\usepackage{bm}
\usepackage{hyperref}
\usepackage{multirow}
\usepackage{array}
\usepackage{booktabs}
\usepackage{ctable}
\usepackage{upgreek}
\usepackage{epsfig}
\usepackage{mathrsfs}
\usepackage{amssymb}
\usepackage{amsbsy}
\usepackage{color}
\usepackage{cancel}
\usepackage{pifont}
\usepackage{marginnote}
\usepackage{float}
\newcommand{\bcen}{\begin{center}}
\newcommand{\ecen}{\end{center}}
\newcommand{\btab}{\begin{tabular}}
\newcommand{\etab}{\end{tabular}}
\newcommand{\bdes}{\begin{description}}
\newcommand{\edes}{\end{description}}

\newcommand{\beq}{\begin{equation}}
\newcommand{\eeq}{\end{equation}}
\newcommand{\bea}{\begin{eqnarray}}
\newcommand{\eea}{\end{eqnarray}}

\newcommand{\bary}{\begin{array}}
\newcommand{\eary}{\end{array}}
\newcommand{\benum}{\begin{enumerate}}
\newcommand{\eenum}{\end{enumerate}}
\newcommand{\bitem}{\begin{itemize}}
\newcommand{\eitem}{\end{itemize}}

\newcommand{\bd} { \mbox{\boldmath $d$}}

\newcommand{\br} { \boldsymbol{r}}

\newcommand{\bA} { \mbox{\boldmath $A$}}
\newcommand{\bB} { \mbox{\boldmath $B$}}

\newcommand{\eqn}[1] {eqn.~(\ref{#1})}

\makeatletter

\newcommand{\Rmnum}[1]{\expandafter\@slowromancap\romannumeral #1@}
\makeatother

\newcommand{\mylabel}[1]{\label{#1}} 

\newcommand{\titlename}{Killing the Hofstadter butterfly, one bond at a time}

\begin{document}



\title{\titlename}
\author{Adhip Agarwala}
\institute{Centre for Condensed Matter Theory, Department of Physics, Indian Institute of Science, Bangalore 560 012, India \email{adhip@physics.iisc.ernet.in}}




\abstract{
Electronic bands in a square lattice when subjected to a perpendicular magnetic field form the Hofstadter butterfly pattern. We study the evolution of this pattern as a function of bond percolation disorder (removal or dilution of lattice bonds). With increasing concentration of the bonds removed, the butterfly pattern gets smoothly decimated. However, in this process of decimation, bands develop interesting characteristics and features. For example, in the high disorder limit, some butterfly-like pattern still persists even as most of the states are localized. We also analyze, in the low disorder limit, the effect of percolation on wavefunctions (using inverse participation ratios) and on band gaps in the spectrum. We explain and provide the reasons behind many of the key features in our results by analyzing small clusters and finite size rings. Furthermore, we study the effect of bond dilution on transverse conductivity($\sigma_{xy}$). We show that starting from the clean limit, increasing disorder reduces $\sigma_{xy}$ to zero, even though the strength of percolation is smaller than the classical percolation threshold. This shows that the system undergoes a direct transition from a integer quantum Hall state to a localized Anderson insulator beyond a critical value of bond dilution. We further find that the energy bands close to the band edge are more stable to disorder than at the band center. To arrive at these results we use the coupling matrix approach to calculate Chern numbers for disordered systems. We point out the relevance of these results to signatures in magneto-oscillations.
}

\maketitle



\section{Introduction}
\mylabel{sec:Intro}

Understanding the role of disorder on electronic conduction has been a central theme in all of  condensed matter physics \cite{Lee_RMP_1985, Kramer_RPP_1993, Janssen_PR_1997,Evers_RMP_2008}.  Apart form being fundamentally interesting from a  theoretical perspective, these problems hold immense significance as they directly bring out (or hide) novel physics in various experimental systems\cite{Abrahams_RMP_2001}. A major milestone in this pursuit has been the scaling theory of localization which stated that any infinitesimal amount of disorder will inhibit any conductivity in a thermodynamically large $2D$ system \cite{Anderson_PR_1958, Abrahams_PRL_1979}. However, a comprehensive understanding of transport in $2D$ is far from complete -- two dimensional systems continue to spring surprises with various phenomena, where mesoscopic physics, interactions, disorder and topology interplay \cite{Stormer_RMP_1999, Novoselov_RMP_2011,Sarma_RMP_2011, Hasan_RMP_2010}.

One of the essential probes in condensed matter is the magnetic field. Effects of which on a $2D$ electron gas leads to integer and fractional Hall effect \cite{Klitzing_PRL_1980, Stormer_RMP_1999}. The same phenomena on an idealized square lattice leads to the Hofstadter model \cite{Hofstadter_PRB_1976}. Interestingly this physics has now been realized both in cold-atomic systems \cite{Aidelsburger_PRL_2013, Miyake_PRL_2013} and material systems \cite{Hunt_Science_2013, Yu_NatPhys_2014}. These systems also possess non-trivial topology and their signatures in transport \cite{Thouless_PRL_1982, Osadchy_JMP_2001}. Not surprisingly, the effect of disorder on quantum Hall physics has received its due attention \cite{Chalker_JPC_1988, Cain_PRB_2001, Galstyan_PRB_1997, Kramer_PR_2005}. For the continuum model -- this question can be posed in two ways -- how does the conductance change when, while keeping the magnetic field same, the disorder is increased or; keeping the disorder same, the magnetic field is reduced. The evolution of the  Landau levels, in a 2D electron gas, and in the lattice setting with a weakening magnetic field has been a matter of debate\cite{Huckestein_RMP_1995, Ortuno_PRB_2011}.  For a recent review refer to \cite{Dolgopolov_PHU_2014}. It was earlier suggested that to be consistent with the scaling hypothesis \cite{Abrahams_PRL_1979}, the Landau levels will float up to higher energies with decreasing magnetic field or increasing disorder \cite{Khmelnitskii_PLA_1984,Laughlin_PRL_1984,Yang_PRL_1996}. However some numerical calculations have hinted otherwise and have instead suggested that the system undergoes a Chern insulator to normal insulator transition as a function of the strength of the disorder \cite{Sheng_PRL_1997}. A two parameter scaling theory has been suggested to understand this transition and a phase diagram was also proposed \cite{Pruisken_PRB_1985,Sheng_PRL_1998,Sheng_PRB_2000}. 

However disorder comes in various varieties -- Anderson disorder\cite{Anderson_PR_1958} is the most famous of them all. In this, random onsite potentials are added to each site of the lattice. The other more stronger kind of disorders are the percolation disorders. They come in two varieties -- site and bond. In the former, one randomly removes the sites from the lattice, in the latter, bonds. Till date, quantum site and bond percolation in $2D$ even in absence of a magnetic field is poorly understood and highly debated -- the central question being -- whether the physics here is different from Anderson disorder \cite{Kirkpatrick_RMP_1973, Mookerjee_Book_2009}. In fact delocalization-localization transition has been predicted in $2D$ for site-dilution on square lattices \cite{Koslowski_PRB_1990, Islam_PRE_2008, Gong_PRB_2009, Dillon_EPJB_2014}.  In this work we limit ourselves to the discussion on bond percolation. If we define $p_b$ as the probability of a link being present between two neighbouring sites, then 
in classical bond percolation, percolation threshold occurs at $p_c=1, 0.5$ and 
$0.2488$ for hyper-cubic lattices in dimensions($D$)=$1,2$ and $3$ respectively \cite{stauffer_Book_1991, Isichenko_RMP_1992}. This threshold signifies the point below which there exists no geometrical connecting path between two sides of a lattice. One expects quantum bond percolation threshold $p_q$ to be $>p_c$ since interference effects will tend to further localize the system even if classically a path may exist. Notice that unlike Anderson disorder here exists a natural bound on $p_q$ due to presence of $p_c$. While finite size scaling analysis
shows an existence of a percolation threshold $p_q$ in $3D$ \cite{Soukoulis_PRB_1991}, results for $2D$ are still
not settled \cite{Mookerjee_Book_2009}.  Some of the previous works have predicted non-zero conductance for $p>p_c$ while others have predicted that all states get localized even for infinitesimal disorder  \cite{Odagaki_PRB_1983, Raghavan_PRB_1981, Shapir_PRL_1982, Taylor_JPCM_1989, Soukoulis_PRB_1991}. A study of transport in bond percolating system and its comparison with classical Drude theory expectations have also been performed \cite{Schmidtke_PRE_2014}. Recently, bond (and site) percolation on a honeycomb lattice has received major attention in order to understand the nature of divergence of density of states at $E=0$ \cite{Sanyal_arXiv_2016, Hafner_PRL_2014, Ostrovsky_PRL_2014, Zhu_PL_2016}.

As far as the effect of magnetic field is concerned, most of the studies above \cite{Sheng_PRL_1997,Sheng_PRL_1998,Sheng_PRB_2000} has been performed for diagonal Anderson disorder. 
A study of banded off-diagonal disorder was performed in \cite{Liu_JPCM_2003}.
However the role of percolation disorder on the Hofstadter model has been little investigated. A periodic dependence of $p_q$ was found as a function of magnetic flux in $3D$ while that in $2D$ was also conjectured \cite{Mier_PRL_1986}. Since for bond percolation disorder the exact value of $p_q$ itself is an open question, it is of particular interest to find if there exists a metal insulator transition before we cross the classical percolation threshold in presence of a magnetic field. 

In this paper, our motivation is two fold. The first part involves understanding the effect of bond percolation disorder on the Hofstadter butterfly pattern as a function of $p_b$. We study the model in both high and low concentration of bond dilutions. We find that even at high amount of bond dilution, we have butterfly-like patterns present in the system. We also look at the effect of bond dilution on band gaps and wavefunctions of the system. We provide understanding of the key features of our results from analyzing small clusters and finite size rings. This provides some physical reasoning behind the results and also contrast them from the case of Anderson disorder. The second part involves calculation of the transport quantities ($\sigma_{xy}$), where a numerical study
based on calculation of Chern numbers is performed using coupling matrix approach \cite{Zhang_CPB_2013}. We study the effect of bond
percolation disorder on Hofstadter bands and show that there indeed is a metal insulator transition with decreasing $p_b$ for $p_b>p_c$. We also find that the Chern bands close to the band edges are more stable to disorder than the ones close to band center, which means that it takes higher disorder strength for achieving metal to insulator transition at the edge of the band than at the center. This result in the low-disorder limit is consistent with the findings for the Anderson disorder case \cite{Sheng_PRL_1997}.

We now present the plan of the manuscript. In the next section (\ref{sec:Hofs}) we provide a brief review of the Hofstadter model and present some of the results for finite rings in presence of magnetic field. These will be used later in our study. In the same section we also introduce the percolation problem. Section. \ref{sec:Kill} contains our results and related discussions on the effect of bond percolation disorder on the Hofstadter butterfly. Here we also discuss the effect of bond dilution on band gaps and on wavefunctions using inverse participation ratios (IPRs). Section \ref{sec:trans} contains the essential details about the coupling matrix approach to calculate the Chern number in the presence of disorder and the corresponding results and discussions. In section \ref{sec:Summary} we summarize our results and speculate some future directions.

\section{Formulation and Prelude}
\mylabel{sec:Hofs}


\begin{figure}
\centering
\includegraphics[width=6cm]{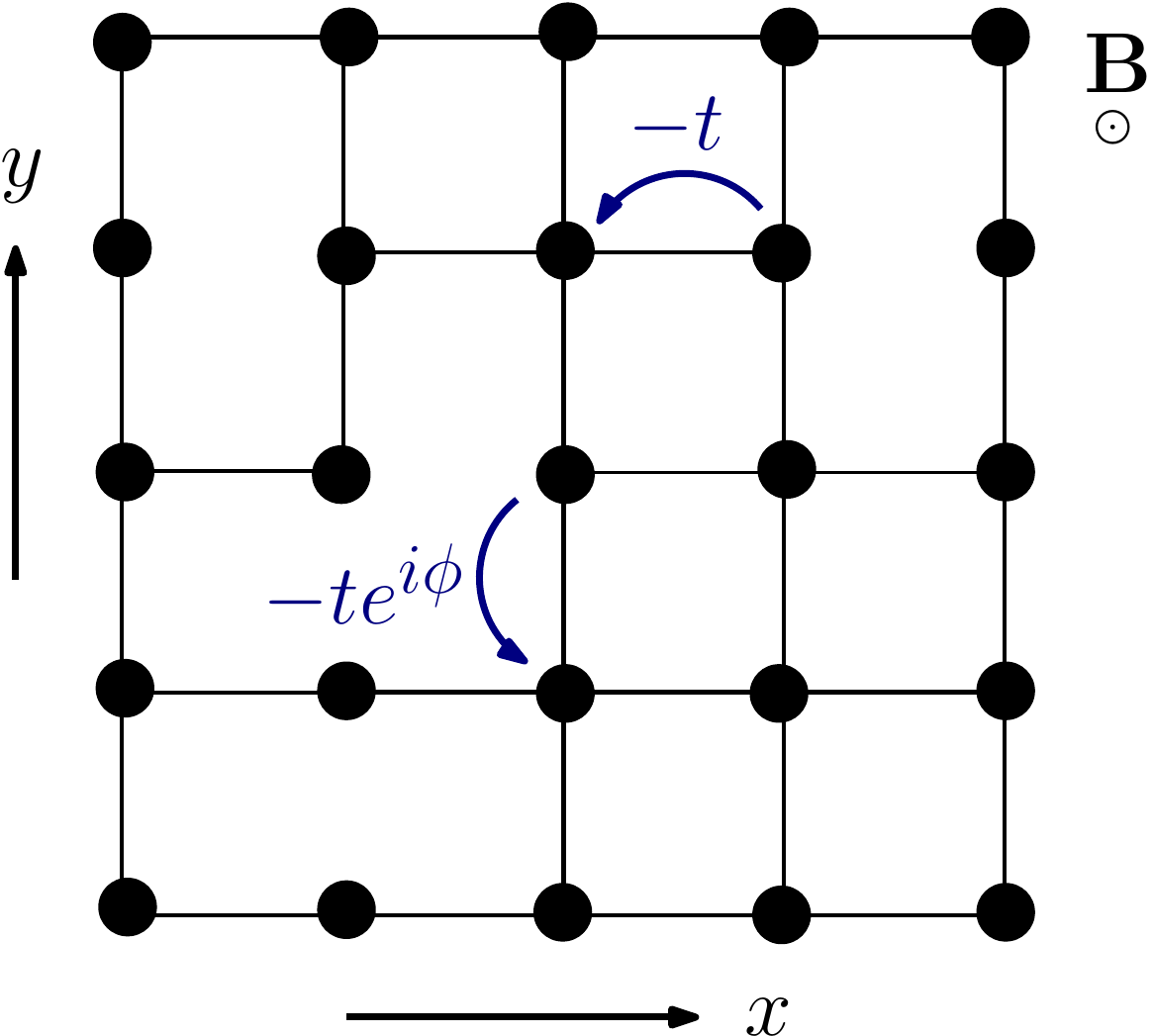}
\caption{(Color online) A schematic of a square lattice with some of the bonds removed. $p_b$ is the probability that a bond is present. Therefore, at $p_b=1$ we have an ideal square lattice. $t$ is the hopping amplitude which is set to 1. $\phi$ is the Pierls phase which includes the effect of the magnetic field {\bf B}.}
\mylabel{fig:schematiclattice}
\end{figure}

\subsection{Hamiltonian}

The Hamiltonian of our interest is, 
\beq
{\cal H} = \sum_{\langle i,j \rangle}-t e^{i\phi_{ij}} c^{\dagger}_{i} c_{j} + h.c.
\eeq

where $c^{\dagger}_{i}$,$c_{j}$ are the creation and annihilation operators for the electrons at site
$i$ and $j$ respectively (see Fig.$\ref{fig:schematiclattice}$). The $\langle i,j \rangle$ signifies that the sum is over the nearest neighbors 
on a square lattice. $\phi_{ij}$ is the Pierls phase which takes into account the effect of a perpendicular
magnetic field $\bB$ on the lattice and is given by 

\beq
\phi_{ij} = \frac{e}{\hbar}\int_{\br_j}^{\br_i}\bA\cdot\bd\br,
\eeq
where $\bA$
is the corresponding vector potential. $t$ is the hopping integral and is set to $1$. $\br_{i(j)}$ denotes the position coordinates of site $i(j)$. We work in Landau gauge where $\bA=(0,Bx,0)$. This conveniently allows for complex phases only in the hoppings in the vertical direction. The flux per plaquette is given by $\alpha$ in units of $h/e$. We will ignore the spin of the fermions. 

\subsection{Hofstadter butterfly}

In the gauge we are using, the system has a translational symmetry in $y$ direction -- therefore $k_y$ is a good quantum number. For a generic $\alpha$, the system does not have translational symmetry in $x$ direction. However, when $\alpha = p/q$, where $p,q$ are integers, the problem can be mapped to a reduced Brillouin zone. The eigenvalues can be plotted as a function of $\alpha$ and this leads to the famous Hofstadter butterfly pattern. This self-similar, fractal pattern was first obtained by Hofstadter \cite{Hofstadter_PRB_1976}, and is reproduced in Fig. \ref{fig:infhofs}.

\begin{figure}
\includegraphics[width=8cm]{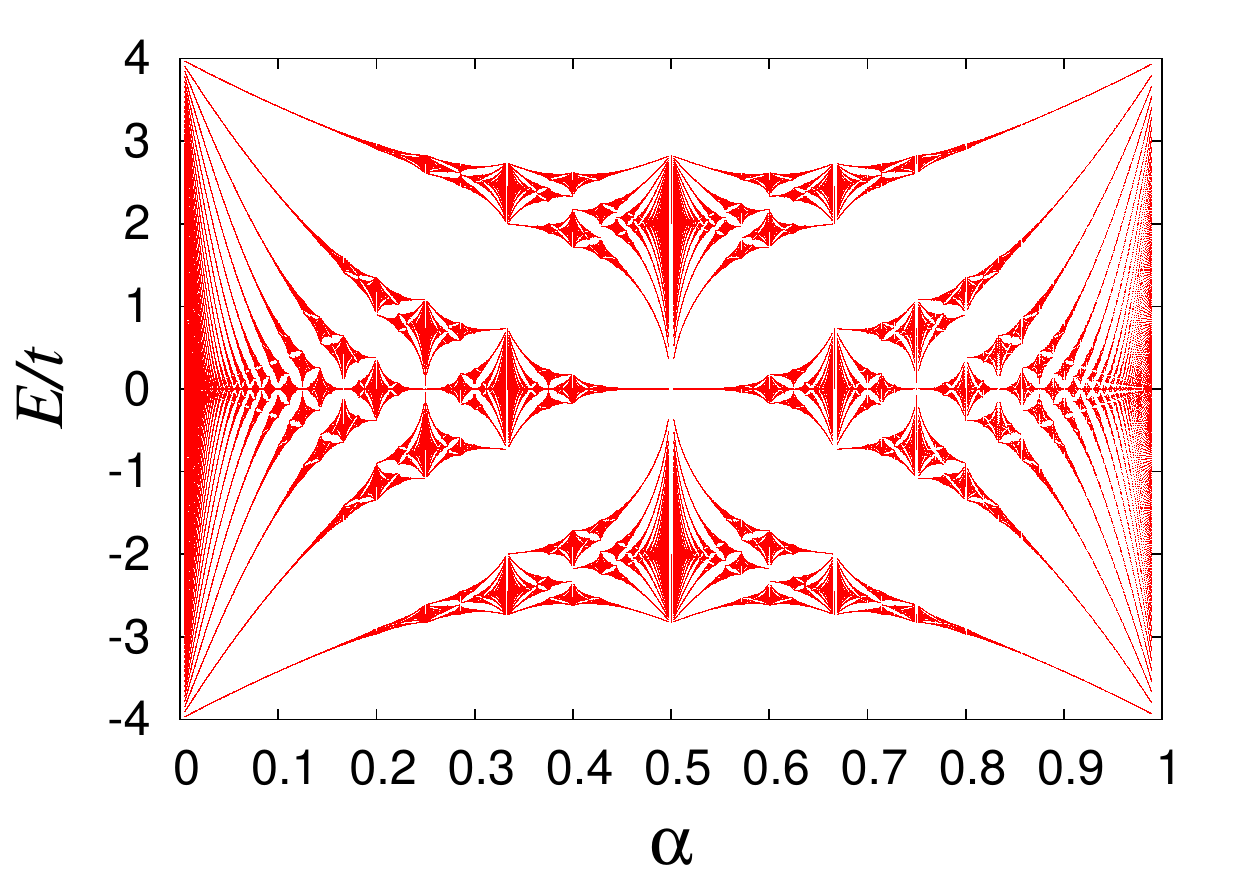}
\caption{(Color online) The plot of the energy dispersion as a function of the 
magnetic flux $\alpha$. This is the Hofstadter butterfly as was originally reported by Hofstadter \cite{Hofstadter_PRB_1976}.}
\mylabel{fig:infhofs}
\end{figure}

\subsection{Polygon in a magnetic field}

Next, let us consider a $N$ sided polygon in a magnetic field. The eigenvalues indexed by $M$ are given by, 

\beq
E_N(M,\alpha_p) = -2 t \cos(\frac{2\pi}{N}(M+ \alpha_p))
\mylabel{polygondis}
\eeq
where $M = \{0,1, \ldots N-1\}$ and $\alpha_p$ is the flux going through the polygon \cite{Analytis_AJP_2004}. Note that $\alpha_p$ is different from the flux per unit plaquette $\alpha$ as introduced in the previous subsection.
Fig. \ref{fig:finhofs} shows the dispersion for few representative finite size rings in presence of uniform magnetic field. As can be seen from Fig. \ref{fig:finhofs}(c) and (d), both the polygons have 8 sides, but the total flux inside the loops are different. While (c) has $\alpha_p = 3 \alpha$, the latter (d) has $\alpha_p = 4\alpha$. These lead to different dispersions ((g)-(h)). These as we will see later will be useful in understanding the results in presence of percolation later.

\begin{figure*}
\centering
\includegraphics[width=14cm]{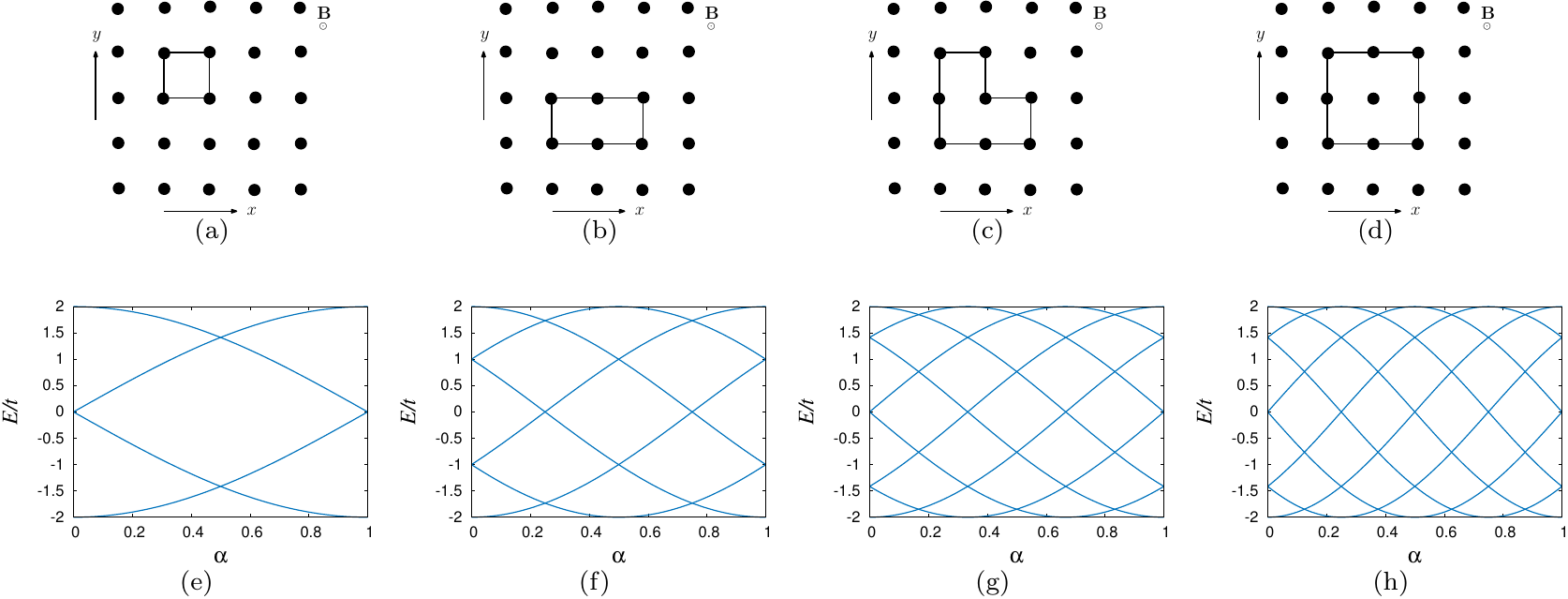}
\caption{(Color online) (a) A connected square with a unit flux $\alpha$ per plaquette has a dispersion shown in (e). (b) A polygon with 6 sides, has two unit squares inside, this corresponds to $\alpha_p =2 \alpha$ in \eqn
{polygondis} and has a dispersion in (f). A 8-sided polygon can have $\alpha_p=3\alpha$(shown in (c)) and $\alpha_p=4\alpha$ as shown in (d). The corresponding dispersions are shown in (g) and (h) respectively.}
\mylabel{fig:finhofs}
\end{figure*}

\subsection{Disorder and Percolation}

Next we define what we precisely mean by percolating the lattice. $p_b$ is defined as the probability of a link being present between two neighboring sites. This implies that at $p_b=1$ we have an ideal square lattice. For any value of $p_b<1$ some of the bonds are removed from the lattice (see Fig.\ref{fig:schematiclattice}). For a square lattice their is a classical percolation threshold at $p^c_b \equiv p_c =\frac{1}{2}$. At any value of $p_b<p_c$ there does not exist a classical geometrical path connecting the two sides of the square lattice \cite{Mookerjee_Book_2009}. Percolation transitions have their own universality classes and distinct critical exponents \cite{Isichenko_RMP_1992}. Percolation is therefore, a special kind of disorder. Even in quantum transport, note that each bond removal is of the energy scale $t$ which is of the same order as the band-width. However density of bonds removed, quantified as $(1-p_b)$, is considered as the tuning parameter of disorder strength. 

Another kind of percolation problem is the site percolation problem. Here sites are randomly removed from a lattice. We state that although, both bond and site percolation problems retain the sublattice symmetry, the bond problem has some `nicer' features than the site percolation. Once a site is removed from a lattice, it effectively reduces the Hilbert space of the problem. Given a imbalance between the number of sites belonging to the two sublattices, one finds $zero$ energy modes in the system, which need to be removed `by-hand' to keep track of non-trivial zero modes \cite{Sanyal_arXiv_2016, Weik_arXiv_2016}. On the contrary, removing bonds on the lattice keeps the dimension of Hilbert space same and only modifies the connectivity between the sites. 

As was mentioned in the introductory section, the most well studied disorder problem is the Anderson disorder \cite{Anderson_PR_1958}. Here onsite potentials to each site is chosen randomly from a distribution (mostly `box') between $[-\frac{W}{t}, \frac{W}{t}]$. Thus $W$ is the parameter characterizing the strength of disorder. A review of numerical results on this can be found in \cite{Markos_APS_2006}.

\section{Killing the Butterfly}
\mylabel{sec:Kill}

In Fig. \ref{fig:bondpercorep} the evolution of the Hofstadter butterfly as function of $p_b$ for some representative values of $p_b$ is shown. While $1-p_b$ can be considered as the `strength' of disorder (like $W$ in Anderson disorder case) we will see that both these disorders are quite different in high disorder limit. Let us first look at the case when $p_b$ is very small and $p_b \ll p_c$ (high disorder limit).

\subsubsection{$p_b \ll p_c$}

In this limit, the system is below the classical percolation threshold and therefore the lattice has already geometrically broken up into disconnected fragments. As can be seen from Fig.\ref{fig:bondpercorep} (j) and (l) one finds that there are many bands which do not disperse with $\alpha$. This can be understood from the fact that most of these structures do not have closed loops which have any magnetic flux passing through. 

We also see that the energies cluster around specific values. To understand these, consider a single site $(j=1)$ connected to $N$ $(j=2,\ldots,N+1)$ other sites with an equal hopping strength $-t$ and no other site is connected to any other. Let the eigenvalues be $\varepsilon_i$, where $(i \in 1,\ldots N+1)$. For a generic $N$ , one finds only two non-zero eigenvalues given by $\pm \sqrt{N}t$. The corresponding eigenvectors are $\frac{1}{\sqrt{2}}(\mp 1, \underbrace{\frac{1}{\sqrt{N}}, \ldots, \frac{1}{\sqrt{N}}}_N)^T$. The other eigenvectors corresponding to $zero$ eigenvalues are of the form $\frac{1}{\sqrt{2}}(0, 0,.., \underbrace{1}_i, .., \underbrace{-1}_j, \ldots 0)^T$, where $i, j$ denotes the site index and take the values $\in (2, \ldots,N+1)$ and therefore has $N-1$ solutions. Since the maximum coordination number for a square lattice problem is $4$, the corresponding non-zero eigenvalues are  $\pm t(N=1)$, $\pm\sqrt{2}t (N=2)$, $\pm \sqrt{3}t (N=3)$ and $\pm 2t(N=4)$. Note that all of these structure have no loops and therefore, the eigen-energies will not change with $\alpha$. Therefore at low $p_b$ limit, as shown in Fig. 4(l), the system has no closed loops and breaks into disconnected fragments. The probability of these structures appearing are $\propto p_b^N$ \cite{Isichenko_RMP_1992}. This therefore also implies that in this limit we have segregation of eigenvalues at some set of discrete energies and DOS peaks only at these specific energy eigenvalues. 

Note that this limit of the Hofstadter model in presence of bond percolation is absolutely distinct from Anderson disorder. The connectivity of each lattice point to the other is not changed in the case of Anderson disorder, and therefore at no value of $W$ do we expect non-dispersing eigenvalues (with $\alpha$). Similarly, increase in $W$ will never lead the eigenvalues to segregate at select eigenvalues. On the other hand, in bond percolation, at $p_b=0$ the DOS will show a $\delta$ function peak at $E=0$. However, these states are distinct from the weak disorder $E=0$ states as also discussed in \cite{Ziman_PRB_1982}, but rather strongly localized states on individual sites which have very high IPR as will be discussed in detail later.

\begin{figure}
\includegraphics[height=20cm]{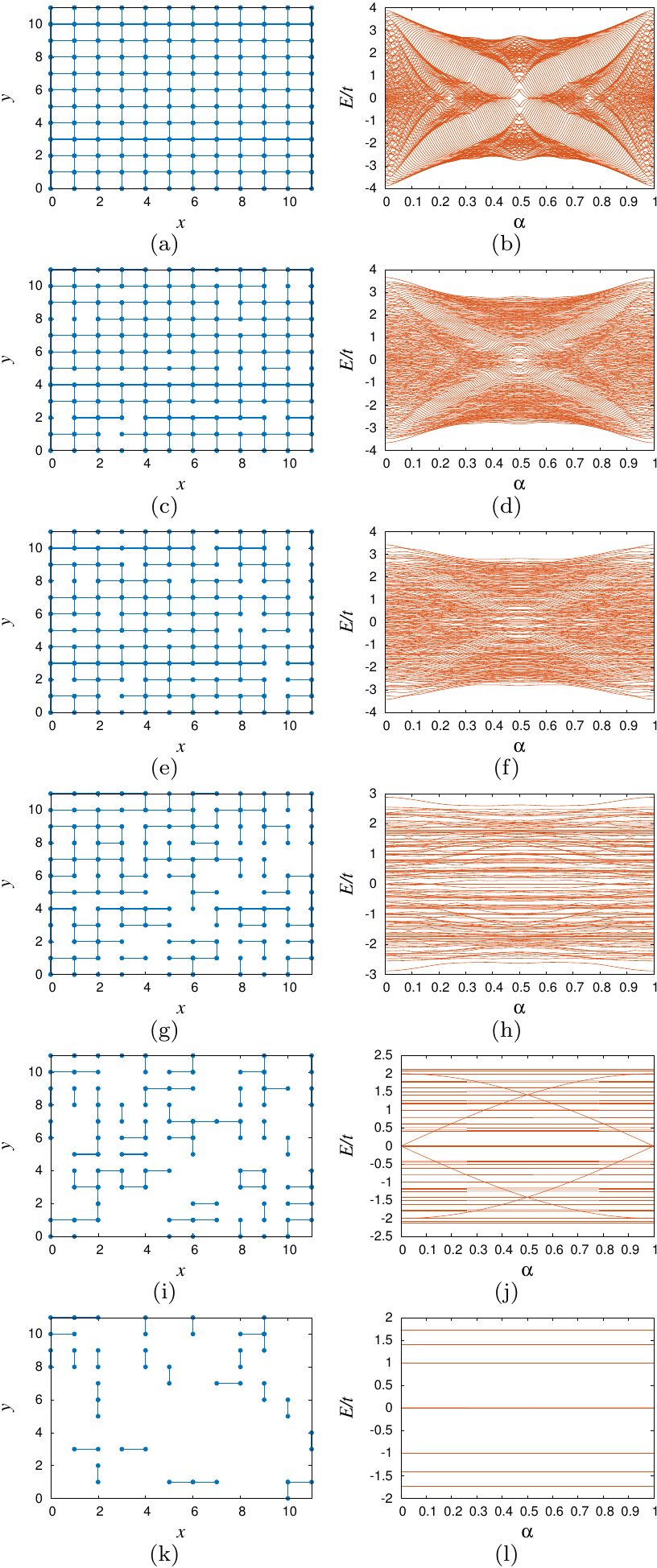}
\caption{(Color online) The representative lattices of $12 \times 12$ size at different values of $p_b$ and their energy dispersion as a function of the magnetic flux $\alpha$. The lattices shown in (a) belongs to $p_b=1$. For (c) $p_b = 0.9$, (e) $p_b=0.75$, (g) $p_b = 0.50$, (i) $p_b=0.25$ and (k) $p_b=0.1$. The corresponding dispersion as a function of $\alpha$ is shown in (b),(d),(f),(h),(j) and (k) respectively. (b) is the corresponding Hofstadter butterfly for a finite $12\times 12$ system. While the increasing disorder destroys the butterfly pattern, one finds non-varying lines present in the dispersion. In (j) one finds only two bands dispersing. While in (l) one finds no dispersing bands.}
\mylabel{fig:bondpercorep}
\end{figure}

\subsubsection{$p_b < p_c$}

As $p_b$ is slightly increased, as can be seen in Fig. \ref{fig:bondpercorep} (j) and (h), dispersing bands appear. While the complete lattice still does-not have a spanning cluster, what is clear is that we have states in the system which disperse with magnetic flux $\alpha$. These are due to small clusters which contain closed loops. Take for example the representative plot shown in Fig. \ref{fig:bondpercorep} (j) and compare the dispersing curve with the Fig. \ref{fig:finhofs} (e). As can be clearly seen they are exactly the same. Therefore, the low $p_b$ ``Hofstadter butterfly'' will be dominated only by these finite size small loops, as shown in  Fig. \ref{fig:finhofs}. These have implications for oscillations in magnetization which we will discuss in more detail later. Note that all these states cannot contribute to transport since they reside only on small clusters.

\subsubsection{$p_b = 1$}
We now discuss the other limit i.e. the clean system. Clearly, even the finite Hofstadter butterfly as shown in Fig. \ref{fig:bondpercorep}(b) has some semblance to the infinite Hofstadter butterfly as shown in Fig. \ref{fig:infhofs}, increasing lattice size makes this similarity more and more apparent \cite{Analytis_AJP_2004}. However, even the finite lattice system has some interesting gap structure at $E=0$ which we now discuss (see Fig. \ref{fig:finhofsE0}). Any finite size square lattice of dimensions ($L\times L$) shows a number of bands dispersing linearly from $E=0$. This number and the slope increases in an interesting fashion, which can be guessed from our discussions on the finite size polygons in magnetic field. Note that a $L \times L$ square ring has $(L-1)^2\alpha$ flux passing through it. Substituting $N=4L-4$, $M=N/4$ in Eqn. \ref{polygondis}, we find the low energy dispersion of the form, 

\bea
 &=& -2t\cos\left(\frac{2\pi}{N}((L-1) + (L-1)^2\alpha)\right) \\ &\approx & (L-1)\pi \alpha
\eea

Now a square lattice of $L \times L$ can contain states on concentric square rings of dimensions $2,4, \ldots, L$, where the states on this rings have small $\alpha$ dispersion as $\pi \alpha,3 \pi \alpha \ldots (L-1)\pi\alpha$ near $E=0$. This can be clearly seen from Fig. \ref{fig:finhofsE0}. As is expected the states indeed lie predominantly on the concentric rings. 

\begin{figure}
\centering
\includegraphics[width=8cm]{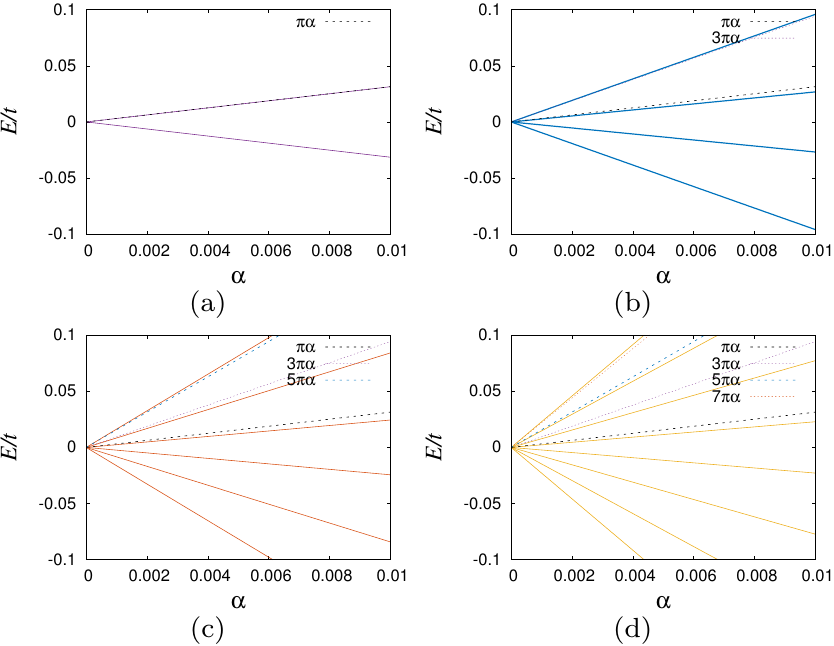}
\caption{(Color online) Band dispersion for square lattices with open periodic condition for lattices of size $L\times L$ for (a) $2 \times 2$ (b) $4 \times 4$ (c) $6 \times 6$ and (d) $8 \times 8$ for $\alpha$ and $E$ close to $zero$. The slope of dispersion approximately follows the slope of $\pi (L-1) \alpha$ }
\mylabel{fig:finhofsE0}
\end{figure}

\subsubsection{$p_b \gg p_c$}

We now look at the effect of the low bond disorder on the finite size Hofstadter butterfly. We focus on the band gap structure at $E=0$. We see from Fig.\ref{fig:finbandgap}, a pure $4\times4$ and $12\times 12$ ((a) and (b)) lattice size has a set of gapless points and large band gaps at some other values of $\alpha$. Increasing disorder, opens up gaps at the gapless points and reduces otherwise large band gaps. This, in some sense, is the usual effect of any disorder i.e. spreading of the $DOS$. It is also reasonable to see that this effect increases with increasing disorder. This is more clear from the inset in the Fig.\ref{fig:finbandgap} where the variance of the gap is plotted. We also note that the amount of gap opened up at $\alpha=0$ is much smaller than other gap-less points. 

\begin{figure}
\centering
\includegraphics[width=8cm]{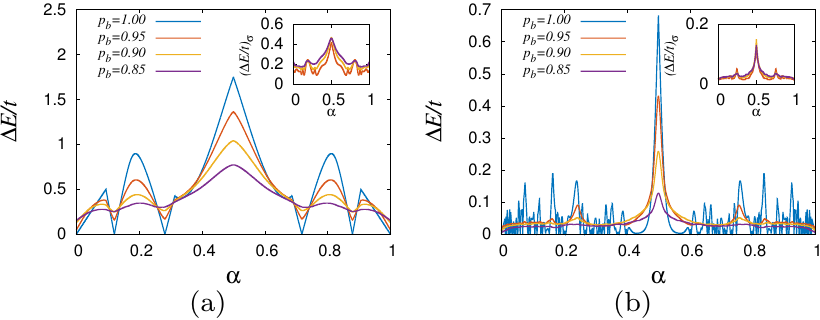} 
\caption{(Color online) The band gap at $E=0$ for four values of $p_b = 1.0, 0.95, 0.90$ and $0.85$ as a function of $\alpha$ for a (a) $4\times4$ lattice and (b) $12 \times 12$ lattice. For the later three values of $p_b$, averages are being plotted over $400$ configurations. The $blue$ line is for the pure system, and has many gapless points and large band gaps including one at $\alpha=0.5$. Increasing the disorder, opens up the gap at gapless points and reduces the magnitude of the larger gaps from the pure system. The lower the value of $p_b$, the effect is larger. In the Inset, the variance ($\equiv (\Delta E/t)_\sigma$) for the later three values of $p_b$ are plotted. It shows the variance $\Delta E/t$ increase with decrease in $p_b$. All these are consistent with our understanding that generic weak disorder spreads out the $DOS$ and open up gaps at the gapless points.}
\mylabel{fig:finbandgap}
\end{figure}

To further understand the effect of bond percolation disorder, we study the Inverse Participation Ratio (IPR) of the different wavefunctions. IPR for a unit normalized wavefunction $|\psi \rangle$ expandable in site basis as, 
\beq
| \psi \rangle = \sum_i \psi_i |i \rangle
\eeq
is given by, 
\beq
IPR = \sum_i |\psi_i|^4
\eeq

This value estimates the spread of a wavefunction in real space. For a delocalized wavefunction spread uniformly over area $A$, IPR $\propto 1/A$, and will decrease with increasing area. If a wavefunction is localized over some few sites, then IPR $\propto 0.1-1$ and doesn't change significantly with increasing size of the system. This diagnostic therefore provides a scope to demarcate localized and delocalized states. To estimate the effect on IPR, we consider 400 configurations of the lattice at a given $p_b$. For each configuration we diagonalize the Hamiltonian and innumerate the energy eigenvalues as $n=1 \ldots L^2$. For each value of $n$ we average over the eigenvalues to find the average energy, and their IPRs to find the average IPR. This gives us the average IPR of the full system as a function of energy and is shown in Fig.  \ref{fig:IPR}.  As the $p_b$ is reduced, IPR at certain values of $E$ becomes very large. Interestingly the values are at $\pm t$, $\pm \sqrt{2}t$ and $\pm \sqrt{3}t$. These correspond to clusters of small sites mentioned before. The corresponding IPRs for these is $1,1/2$ and $1/3$. As $p_b$ is decreased further some of these peaks vanish, and now only the central peak remains. Also peaks at other values correspond to the solutions for a open tight binding chain. For a $n$-site chain the dispersion is given by $-2t\cos k$, where $k \in \frac{m\pi}{n+1}$, where $m \in (1,2\ldots n)$. For example, a 4-site open chain has eigenvalues at $\pm 2\cos(\frac{\pi}{5}) (\sim \pm0.62)$ and $\pm 2\cos(\frac{2\pi}{5}) (\sim \pm 1.62)$ which can also be clearly seen in Fig. \ref{fig:IPR} (a).

In Fig. \ref{fig:IPR}(b) we look at relatively smaller values of $p_b$ and look at the effect of increasing the system size. The variation of IPR signifies whether the system is comprised of localized or delocalized states. For example at $p_b=1.0$ one finds that IPR is quite low  ($\sim 5 \times 10^{-3}$) and reduces with increasing system size. Decreasing $p_b$ one notices a strong peak appears at $E=0$, implying appearance of localized states. However, one notices that decreasing $p_b$ more high IPR peaks start to appear at distinct energy values as discussed in detail above. However, interestingly the average IPR of the system increases, and the value for $30\times30$ starts to overlap with $24\times24$, implying localization overall in the full spectrum. This feature becomes quite prominent at $p_b \lesssim 0.65$. Note that the average IPR is about $0.025$ at $p_b=0.60$ signaling that the wavefunction resides only on a average of 40 sites, in otherwise a lattice of 900 sites. This signals the wavefunctions have got localized much before the classical percolation threshold is reached. This will be investigated more clearly through calculations of the transport in the next section.

\begin{figure}
\centering
\includegraphics[width=7cm]{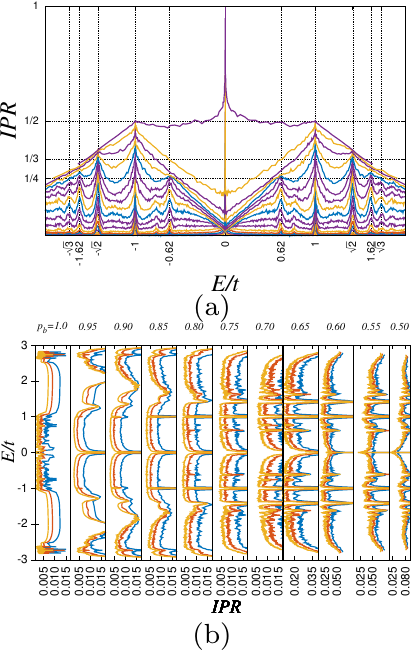} 
\caption{(Color online) Variation of IPR as a function of bond probability. (a) IPR is shown for a $30 \times 30$ lattice for $p_b$ starting from $1.0$ (bottom-most) to $0.05$(topmost) in intervals of $0.05$. One notices appearance of peaks at specific values of $E/t$ which goes away with reducing $p_b$. (see text)(b) IPR for system sizes $18\times18$ (blue), $24 \times 24$ (orange) and $30 \times 30$ (yellow) compared to each other at $\alpha = 1/4$ when averaged over $400$ configurations. At $p_b=1.0$ the IPR is quite small ($\sim 5 \times 10^{-3}$) and reduces with increasing system size. Decreasing $p_b$ one notices a strong peak appears at $E=0$, implying appearance of localized states. However, one notices that decreasing $p_b$ more high IPR peaks start to appear at distinct energy values. Also the average IPR of the system increases for a $30\times30$ lattice and overlaps with the case of $24\times24$, implying overall localization in the spectrum. This feature becomes prominent for $p_b \lesssim 0.65$. Error bars are not shown for clarity of figure.}
\mylabel{fig:IPR}
\end{figure}

\section{Effect on Chern numbers and Transport}
\mylabel{sec:trans}

Hofstadter Model, apart from structure of the eigenspectrum, also hosts interesting structure of the topological invariants \cite{Thouless_PRL_1982, Osadchy_JMP_2001, Fradkin_Book_1991}. It will be interesting to understand the effect of disorder on such topological invariants. This therefore requires calculation of Chern numbers. Note that in presence of disorder, the system no longer contains translational symmetry and therefore a momentum integral over the Brillouin zone will not suffice to calculate the Chern number. We therefore calculate the Chern numbers using the method outlined in \cite{Zhang_CPB_2013}. This essential numerical technique is motivated from the fact that Chern number can also be calculated from an integral over the twisted boundary conditions. For completeness we 
include briefly some of the definitions and a brief discussion about the method following \cite{Zhang_CPB_2013}.

 For a $2D$ lattice comprising of $N=L\times L$ unit cells,
the single particle wavefunctions can satisfy the following boundary conditions given by
$\phi_{\theta}(x+L,y)=e^{i\theta_x}\phi_{\theta}(x,y)$ and $\phi_{\theta}(x,y+L)=e^{i\theta_y}\phi_{\theta}(x,y)$,
where $\theta = (\theta_x,\theta_y)$ such that $0 \le \theta_x,\theta_y \le 2 \pi$.  
For a given filling, we can have $M$ states occupied. Let the many body wavefunction of these $M$ states be written as $\Psi_{\theta}$.
Then the Chern number of the ground state is given by 
\beq
C = \frac{1}{2\pi i}\int_{T_\theta} d\theta \langle \nabla_{\theta} \Psi_{\theta}|\times| \nabla_{\theta} \Psi_{\theta} \rangle
\eeq
where $T_{\theta}$ denotes the allowed $(\theta_x, \theta_y)$ values \cite{Niu_PRB_1985}. 

Note that since in defining $\Psi$ we have taken into
consideration all the filled states, $C$ here is the sum of the Chern numbers of individual bands below the chemical
potential. Hence the quantity evaluated can be interpreted as $\sigma_{xy}$ in units of $e^2/h$. 

The calculation of $\sigma_{xy}$ can be directly done using Lehmann representation of the Kubo formula \cite{Dutta_JAP_2012}.  However finding the Chern number of each band requires numerical diagonalization of a system many number of times\cite{Sheng_PRL_1997}. The recently developed coupling matrix approach allows for a much simpler and numerically inexpensive method \cite{Zhang_CPB_2013}. The idea is to convert the integral over $T_\theta$ into an integral over a path in momentum space. This integral then can be solved as a product of matrices whose components are determined by the inner product of some wavefunctions which were determined only by the system under periodic boundary conditions. The essential simplifying step is to do away with the necessity of diagonalizing the system at different values of boundary conditions. This approach can also take into consideration the effect of real space disorder in a natural way.

Now we present the results of our calculations.
In Fig.\ref{fig:flux1by4finsz} we show the variation of $\sigma_{xy}$ with bond occupation probability $p_b$ for $p/q=1/4$.
The filling is kept constant at $1/4$ and for each configuration chemical potential is self consistently evaluated.
The lattice size is systematically increased from $12\times 12$ to $24 \times 24$ in difference of $4$ sites per side.
We find that increasing lattice size makes the transition sharper and clearly the conductance goes to zero much before $p_c$ which
occurs at $p_b=0.5$. The transition seems to occur close to $p_b \approx 0.65$, which was also tentatively the value seen from IPR results in previous section.

\begin{figure}[h!]
\includegraphics[width=8cm]{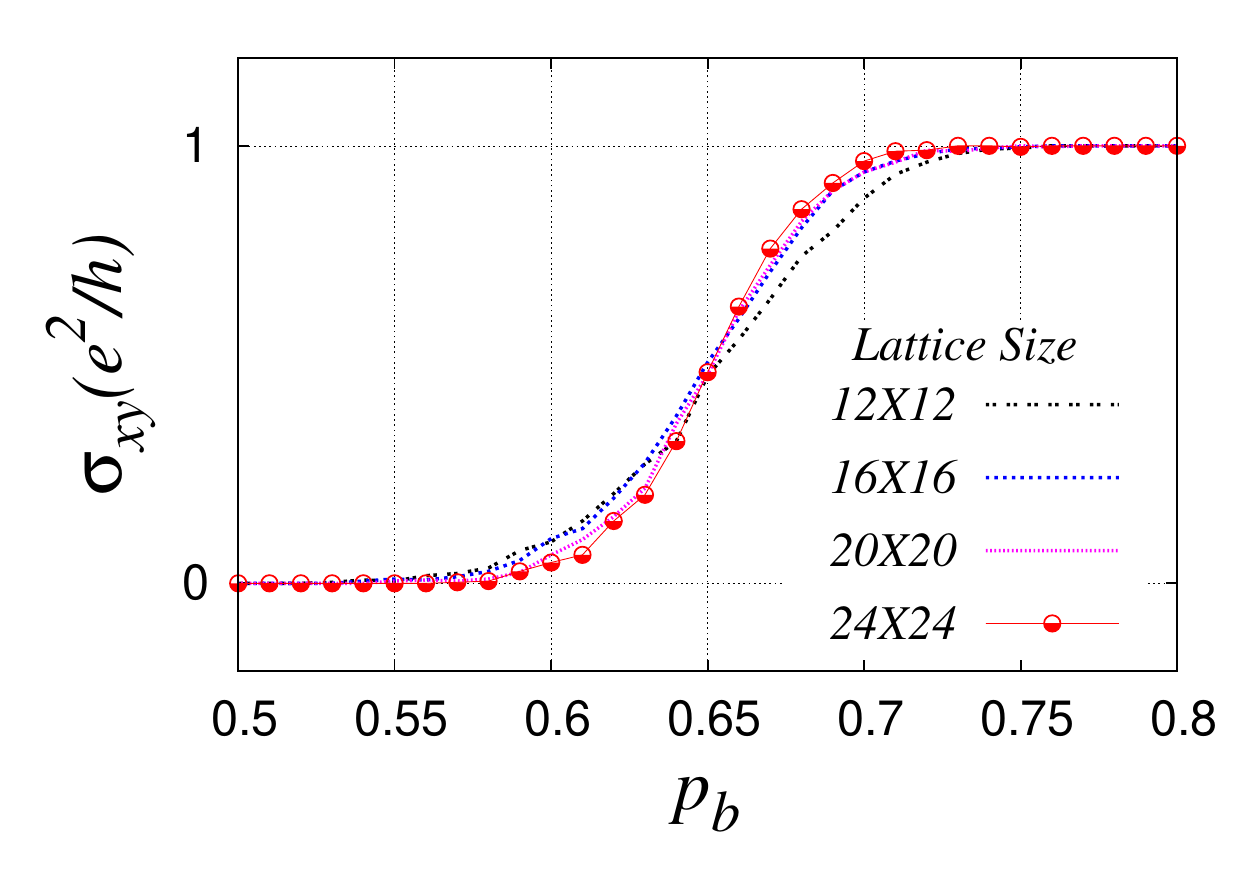}
\caption{(Color online) Plot of $\sigma_{xy}$ with bond occupation probability $p_b$ at $p/q=1/4$ for different lattice sizes. The filling is kept constant at $1/4$ and the lattice size is increased from $12\times 12$ to $24 \times 24$. The red line with moon points is for the lattice size of $24\times 24$. The results are averaged over 400 disorder configurations. With increasing lattice size we find the transition becoming sharper around $p_b \approx 0.65$.
The $standard$ $error$ of the mean is of the order of the point size or lower and hence has not been shown above.}
\mylabel{fig:flux1by4finsz}
\end{figure}

In Fig.\ref{fig:flux1by16fill} we show the variation of {$\sigma_{xy}$ with bond occupation probability $p_b$ for
$p/q=1/16$ for different fillings. We find that with increasing filling the Chern
insulator plateau remains stable only for lower strength of bond disorder. Note that with increasing filling
from $1/16$ to $7/16$ we moved from bottom of the spectrum to band center. This resembles what was found for the Anderson
disorder in earlier studies \cite{Sheng_PRL_1997}. 

\begin{figure}[H]
\centering
\includegraphics[width=8cm]{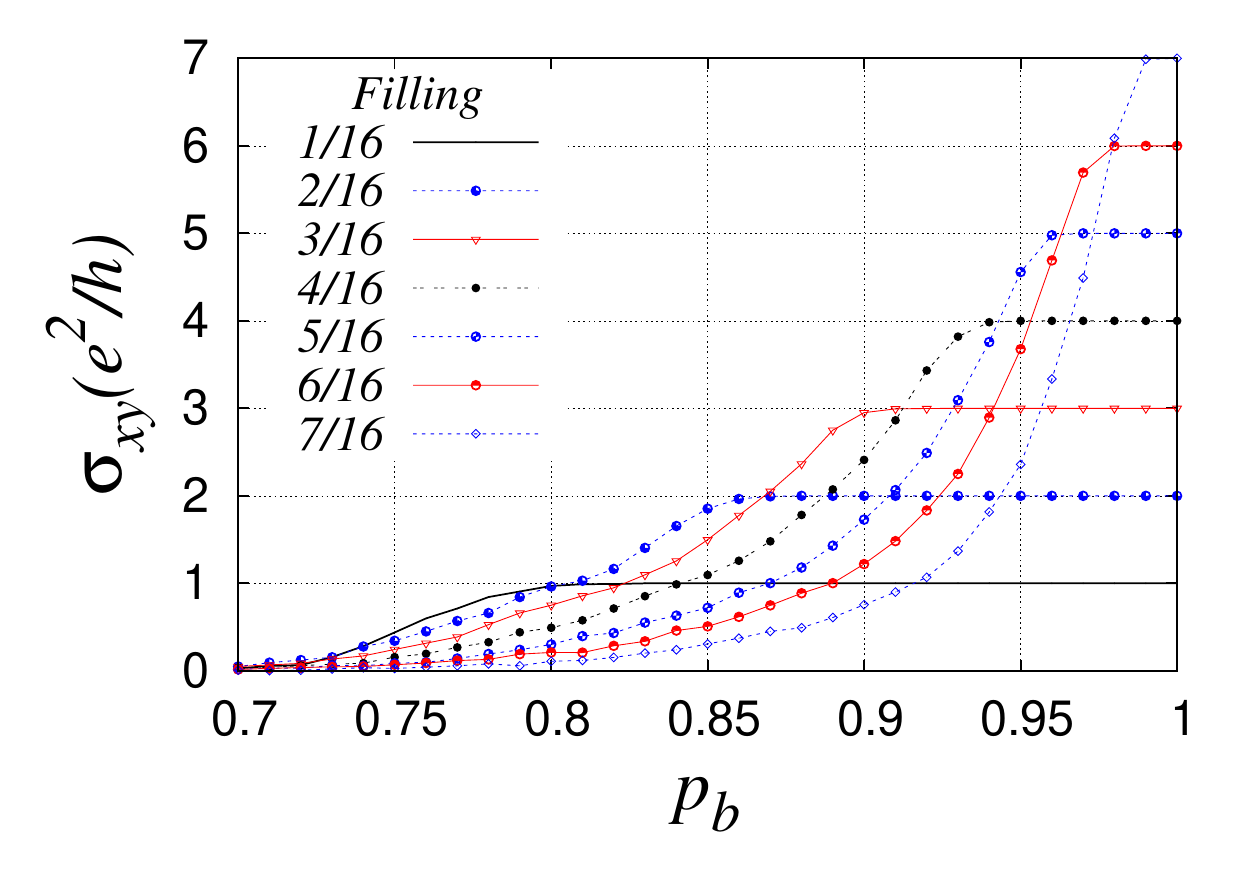}
\caption{ (Color Online) $\sigma_{xy}$ with bond occupation probability $p_b$ for $p/q=1/16$ for different fillings.
The lattice size is $32 \times 32$.
The results are averaged over 400 disorder configurations. In the clean limit ($p_b=1$) $\sigma_{xy}$ for fillings $n/16$ is $n \frac{e^2}{h}$ where ($n \in 1,\ldots,7$). With increasing filling one moves from the bottom of the full spectrum to the center. The Chern insulator plateaus are less stable to bond disorder as one moves closer to band center.}
\mylabel{fig:flux1by16fill}
\end{figure}

To understand the underlying mechanism for this, one first realizes that the physics of the Hofstadter problem is different from that of the continuum $2D$ model. 
Unlike the continuum, $\sigma_{xy}$ can be negative in the lattice setting \cite{Fradkin_Book_1991}. This is because the bands here
may carry negative Chern numbers. For an even $q$ a negative Chern number band of Chern number $-2(q-1)$ lies at the band center. 
With increasing onsite disorder is has been argued that this central band mixes with the other bands hence
explaining why the bands close to band center are the first ones to show transition to normal insulator \cite{Sheng_PRL_1997}. We infer that the similar considerations indeed apply in this limit of bond percolation problem.

 Also there is a lot of interest in understanding the physics 
in the limit of $zero$ magnetic field. We expect that with weakening magnetic field, the amount of bond percolation disorder required for Anderson insulating transition will decrease  i.e. $p_q$ will slowly approach a larger value. This can be expected from Fig. \ref{fig:flux1by4finsz} and Fig.  \ref{fig:flux1by16fill}.  The magnetic flux is kept high $(p/q=1/4)$ in Fig. \ref{fig:flux1by4finsz}   and 
the transition occurs at $p_b \approx 0.65$. While in Fig. \ref{fig:flux1by16fill} the magnetic flux is kept low $(p/q=1/16)$, 
and the transitions for all values of filling occurs at $p_b>0.65$. This suggests that with 
further decrease in magnetic field, one might expect higher values of $p_b$ (lower number of bonds removed) where the transition will occur. The exact form of this variation and its filling dependence would be interesting to investigate.

\section{Summary and Future Directions}
\mylabel{sec:Summary}

To summarize, we have studied the effect of the bond percolation disorder on the Hofstadter bands
which are formed when a perpendicular magnetic field is applied to a square lattice.
We have looked at the evolution of the Hofstadter butterfly as the bond percolation is increased. We find that at low values of $p_b$, unlike the Anderson disorder, the eigenspectrum does-not disperse with the magnetic field. This we attribute to the open clusters of sites which do not enclose any magnetic field. With slight increase in $p_b$ we find few dispersing states which are due to disconnected rings. The dispersion of these are compared with finite size ring structures. At large values of $p_b$ we find the bond percolation spreads out energy eigenvalues and the gapless points get opened up. We also analyze the IPR of the wavefunctions as a function of $p_b$, and have looked at the effect of this disorder on band gaps and states close to $E=0$. To understand some of the features of our results we discussed properties of finite size rings and clusters.

Next we investigated the effect of disorder on the Chern bands, and found that they undergo direct transition to the normal insulator state with increasing bond percolation disorder. This happens at a
 bond occupation probability $p_b$ higher than the classical percolation threshold.
We also find that the bands at the band bottom are more stable to disorder than the band center. The calculations
were performed using a recently developed method of calculating Chern numbers using coupling
matrix approach \cite{Zhang_CPB_2013}. These results seem to be in accordance with the insights found 
from the diagonal Anderson disorder problem \cite{Sheng_PRL_1997}.

We now mention some of the future directions.  In our study, we have looked at two aspects of the physics of bond percolation on square lattices when kept in presence of uniform magnetic field. One, the effect on 
the energy dispersion, which leads to the effective ``killing" of the Hofstadter butterfly. And two, effect on transverse conductivity $\sigma_{xy}$.  It will be interesting to look at the magnetic oscillations in this system for a fixed density of particles. Magnetization($M$) is determined by the change of the energy dispersion of the system as a function of magnetic field $M = -\frac{\partial E}{\partial \alpha}$ \cite{Analytis_SM_2005}. If the energy spectrum does not disperse with magnetic field ($\alpha$), as is the case when $p_b \ll p_c$, then this quantity will be identically $zero$. Therefore absence of oscillations due to increasing bond-percolation is a signature of reaching the limit of high percolation disorder. However, the exact form of this change and the variation with $p_b$ may be interesting to understand.

Further, while we study the effect of bond percolation on the $\sigma_{xy}$, it might be interesting to correlate this with the effect on $\sigma_{xx}$. It will be intriguing to understand if the two-parameter scaling theory, as has been tested for other disorder problems in quantum Hall physics \cite{Sheng_PRL_1998}, is also applicable to the bond percolation disorder. There are other interesting parallels between the Anderson transition and percolation transition which would be worth pursuing. It was shown in \cite{Kaneko_JPSJ_2003}, that both percolation and Anderson transition have a characteristic exponent by which the radius of a wavepacket spreads with time. It might be interesting to study these exponents in presence of a magnetic field and investigate the transitions from the Hall state to an Anderson insulator state. In  \cite{Goldenfeld_PRB_2006} existence of weaker Anderson transitions was shown for diagonal disorder in $2D$. It might also be interesting to realize this physics in case of percolation disorder and study its interplay with the effects of a perpendicular magnetic field.

\begin{acknowledgement}
Financial support from CSIR, India is gratefully acknowledged. AA is grateful to Vijay B. Shenoy for suggestions, discussions and computational resources. AA also thanks Sambuddha Sanyal, Jayantha P. Vyasanakere and Ajit C. Balram for discussions and for comments on the manuscript.
\end{acknowledgement} 

\bibliographystyle{unsrt}
\bibliography{refChern}

\end{document}